\documentstyle[twoside,fleqn,npb,bm,epsfig,amsmath]{article}
\title{Continuum methods in lattice perturbation theory}

\author{Thomas Becher\address{Stanford Linear Accelerator Center, Stanford University,
Stanford, CA 94309, U.S.A. }%
        \thanks{Speaker}
        and
        Kirill Melnikov\address{Department of Physics and Astronomy, Univ.~of Hawaii, 2505 Correa Rd., Honolulu, HI 96822, U.S.A.}}

\begin{document}

\begin{abstract}
We show how methods of continuum perturbation theory can be used to
simplify perturbative lattice calculations. We use the technique of
asymptotic expansions to expand lattice loop integrals around the
continuum limit. After the expansion, all nontrivial dependence on
momenta and masses is encoded in continuum loop integrals and the only
genuine lattice integrals left are tadpole integrals. Using
integration-by-parts relations all of these can be expressed in terms
of a small number of master integrals. Four master integrals are
needed for bosonic one loop integrals, sixteen in QCD with Wilson or
staggered fermions.
\end{abstract}

\maketitle

\noindent\hspace*{-1.9mm} 
\raisebox{8cm}[0ex][0ex]{ 
{\normalsize 
\parbox{4cm}{ 
{\textsf{SLAC-PUB-9568}}
{\textsf{hep-ph/0211215}}}} 
}\vspace*{-5ex}

\section{Introduction}

In many cases, the results of numerical simulations of QCD on the
lattice need to be matched by equally precise {\em perturbative}
calculations in lattice regularization to become phenomenologically
relevant. In the past, the need (or the motivation) for precise
matching calculations has been limited by the large systematic errors
induced by the quenched approximation. By now, simulations with fairly
light dynamical quarks are feasible and precise calculations in
lattice perturbation theory are becoming increasingly important.  The
dominant uncertainty in the first determinations of the strong
coupling constant from lattice simulations with three dynamical
quarks, for example, are unknown higher order perturbative corrections
to the relation between the coupling measured on the lattice and the
$\overline{\mbox{MS}}$ coupling \cite{Davies:2002mv,DiPierro:2002ta}.

The introduction of a space-time lattice to regulate the theory
considerably complicates perturbative calculations. The propagators
and vertices become functions of sines and cosines of the momenta and
not a single loop integral in four dimensions can be evaluated
analytically. In the standard approach to lattice perturbation theory
the relevant diagrams are therefore evaluated numerically, which has
several drawbacks: (a) the amount of numerical computations necessary
for realistic calculations is huge; (b) cancellations between
individual diagrams can render numerical results unstable; (c) the
continuum limit, i.e. the limit in which the inverse lattice spacing
becomes much larger than external momenta and masses, has to be taken
numerically as well. There are a number of techniques to both reduce
the number and increase the precision of the numerical integrations
involved \cite{Caracciolo:1991cp,Burgio:1996ji,Luscher:1995zz}, but we
feel that the computational tools for perturbative calculations on the
lattice are not as highly developed as the methods used in the
continuum.

In a recent paper \cite{Becher:2002if}, we have proposed to use
continuum methods to simplify lattice calculations. In particular, two
tools originally developed for the evaluation of continuum loop
integrals turn out to be useful: the technique of asymptotic
expansions \cite{Smirnov:pj} and the use of integration-by-parts
relations \cite{ibp}. The method of asymptotic expansions can be used
to expand lattice loop integrals around the continuum limit.  This
expansion splits the loop integrals into two parts: the ``hard'' part
is a sum of lattice tadpole integrals, while the ``soft'' part
consists of ordinary continuum integrals. In a second step, we use
integration-by-parts relations to express all lattice tadpole
integrals through a few master integrals. In a theory involving only
bosonic fields, there are four such master integrals at one loop. In
QCD with Wilson or staggered fermions, the number of master integrals
increases to sixteen. The master integrals can be chosen to be
convergent and are evaluated numerically.

We begin by quickly reviewing perturbation theory in lattice
regularization and some of the complications that arise with this
particular regulator. After illustrating the above method with a
simple example, we discuss its application to the case of staggered fermions.

\section{Lattice perturbation theory}

When calculable at all, phenomenological results in QCD are often
obtained in a factorized form: a physical quantity is given as a
product of a perturbative short-distance part times a low-energy
contribution.  The splitting into the two parts is regularization
dependent. If one manages to evaluate the low energy part with a lattice
simulation, one also needs the high energy part in the same
regularization. This requires a perturbative calculation in lattice
regularization. Such calculations are, however,
tedious. First of all, the propagators are much more complicated
than in the continuum. The simplest discretization of the bosonic
propagator in four dimensions is
\begin{equation}
G_B(k) = \frac{1}{(\widehat{k}^2 + m^2)}\, ,
\label{prboson}
\end{equation}
where
\begin{equation}
\widehat{k}^2=\sum_{\mu=1}^4 \widehat{k_\mu}^2~~ \text{ and }~~
\widehat{k_\mu} = 2 \sin\frac{k_\mu}{2}\, .
\end{equation}
Loop integrals involve products of these propagators, integrated over the Brillouin zone; a typical loop integral has the form
\begin{equation}
\int \limits_{-\pi}^{\pi}\frac{d^4 k}{(2\pi)^4}  \frac{1}{(\widehat{k}^2
+ m^2)}\,\frac{1}{((\widehat{p+k})^2 + m^2)}\,,
\label{e1}
\end{equation}
where $p$ is the external momentum. Note that we have expressed all quantities in units of the
lattice spacing $a$: $k_\mu=k^\mu_{\rm phys}\,a$, $m=m_{\rm phys}\,a$, etc.

In lattice gauge theories not only the propagators, but also the
vertices take a rather complicated form. The gluonic action is given
in terms of Wilson loops, whose expansion in terms of the gauge field
yields interactions among any number of gluons. The four-gluon vertex
arising from the simplest Wilson loop, for example, involves one hundred
terms with up to four powers of sines and cosines of the in- and
outgoing momenta (see e.g.~\cite{Rothe:kp}).

\section{Asymptotic expansion around the continuum limit}

We now discuss the expansion of the massive tadpole integral
\begin{equation}
G(m) = \int_{-\pi}^{\pi}\frac{d^4k}{(2\pi)^4}\,\frac{1}{(\widehat{k}^2 + m^2)}
\, ,\label{eq:tad}
\end{equation}
around the continuum limit, corresponding to the expansion around
$m=m_{\text{phys}}\,a=0$. Altough this is the simplest possible
example, it illustrates all the important features of our approach.

We begin by mapping the integration region in Eq.(\ref{e1}) to an
infinite volume and defining new integration variables $\eta_\mu$,
\begin{equation}   
\eta_\mu = \tan(k_\mu/2).
\label{beq}
\end{equation}
In terms of the new variables, the loop integrations in Eq.(\ref{eq:tad})
range from $-\infty$ to $+\infty$:
\begin{equation}
G(m)=\frac{1}{4\pi^4}\!\!\int\limits_{-\infty}^{\infty} \prod_{i=1}^4
\frac{d\eta_i}{(1+\eta_i^2)} \left [ \frac{m^2}{4}+ D_B(\eta) \right ]^{-1-\delta}\!\!,
\label{eq2}
\end{equation}
where 
\begin{equation}
D_B(\eta) = \mbox{$\sum \limits_{i=1}^4$}
\frac{\eta_i^2}{(1+\eta_i^2)}.
\end{equation}
After the variable transformation, the form of the integrand becomes
more similar to a continuum integral. In addition to the change of
variables, we have also introduced an {\em intermediate regulator} by
raising the propagator to the power $1+\delta$ in Eq.~(\ref{eq2}).

The expansion of $G(m)$ is obtained as a sum of two parts:
\begin{equation}
G(m) = G_{\rm soft} (m) + G_{\rm hard}(m).
\label{eq4}
\end{equation}
The soft and hard contributions are calculated by applying
the following procedure to the integrand in Eq.(\ref{eq2}):
\begin{itemize}
\item Soft: assume that all the components of the
loop momentum $\eta$  are small, $\eta_i \sim m \ll 1 $.
Perform the Taylor expansion of  the integrand in Eq.(\ref{eq2}) in
the small quantities $\eta_i$ and $m$. 
The expansion coefficients in this region
are standard continuum one loop integrals, regularized analytically.
Note that {\it no} restriction on the integration region is
introduced.
\item Hard: assume that all the components of the loop momentum are large,
$\eta_i \sim 1 \gg m$ and Taylor expand the integrand in $m$. 
The expansion coefficients are massless lattice tadpole integrals.
\end{itemize}
As in dimensional regularization, scaleless integrals are set to
zero. Higher orders of the soft part contain more and more ultraviolet
divergent integrals. The ultraviolet divergences manifest themselves
as poles in $\delta$:
\begin{multline}
G_{\text{soft}}(m)=\frac{m^2}{16\,{\pi }^2}\,\left( -1 - \frac{1}{\delta }
  +2\ln\frac{m}{2} \right)\\+\frac{m^4}{256\pi^2}\,\left(3 +
  \frac{2}{\delta } - 4\ln\frac{m}{2} \right)+O(m^6)\,.
\end{multline}
In the final result, these poles will cancel against infrared
divergences appearing in the expansion of the hard part, whose
evaluation we discuss in the next section.

\section{Integration-by-parts relations between lattice integrals}

While the soft part is given in terms of continuum integrals, the hard part involves the functions 
\begin{equation}\label{basishard}
H(\{a_i\};n)=  \int \limits_{-\infty}^{\infty}
\prod_{i=1}^4 \frac{d\eta_i}{(1+\eta_i^2)^{a_i}}
\left [D_B(\eta) \right ]^{-n-\delta} \, .
\end{equation}
The functions $H(\{a_i\};n)$, with $a_i$ and $n$ integers, provide the
full set of genuine lattice integrals needed for any one-loop
calculation.  The integrals occuring in the hard part are process
independent and can therefore be calculated once and for all for a
given lattice action (the hard integrals occuring in the case of
staggered fermions are given in section \ref{sec:staggered}).

The hard part of Eq.~(\ref{eq2}) is obtained by expanding the integrand in $m$
\begin{multline*}
4\pi^4\,G(m)_{\text{hard}}=H(\vec{1},1 )- 
 \frac{m^2}{4}\left( 1 + \delta \right) H(\vec 1,2 )\\ +
\frac{m^4}{32}\,\left( 1 + \delta\right)\left(2+\delta \right)\,H(\vec 1,3)+\dots \, ,
\end{multline*}
with $\vec{1}=\{ 1,1,1,1\}$.
The hard part of integrals with external momenta or tensor structure also involves functions $H$ with index values $a_i\neq 1$.

We now use integration-by-parts identities to fully exploit the
algebraic relations between the  various integrals $H(\{a_i\},n)$. 
These relations are derived using the fact that the integral of a total 
derivative vanishes in analytic regularization:
\begin{equation}
0=  \int \limits_{-\infty}^{\infty} d^{d}\eta
\frac{\partial}{\partial \eta_\mu} \Big \{ \eta_\mu
\prod_{i=1}^4 \frac{1}{(1+\eta_i^2)^{a_i}}
\left [ D_B(\eta) \right ]^{-n-\delta}  \Big \} \, ,
\label{recrel}
\end{equation}
for each value of $\mu=1,2,3,4$.  Computing the derivative in
Eq.(\ref{recrel}) and rewriting the resulting expression in terms of
the functions $H(\{a_i\},n)$, we obtain an algebraic relation between
integrals with different values of $\{a_i\}$ and $n$. Another equation
can be obtained by partial fractioning, i.e.~by using the linear
dependence of the five ``propagators'' in the function $H(\{a_i\},n)$.
The complete set of algebraic relations is therefore
\begin{eqnarray}\label{eq:scalarRel}
0 &=& \bigg\{{\bm n}^- + \sum_{i=1}^{d} ({\bm a}_i^+-1)\bigg\}
H\, , \\
0 &=& \bigg \{ 1 + 2\,a_i\, ({\bm a}_i^+ - 1)   
\nonumber \\
&& 
+ 2\,(n+\delta)\,{\bm n}^+\,{\bm
a}_i^{+}({\bm a}_i^{+}-1)\bigg \} H \, .
\nonumber 
\end{eqnarray}
The conventions are such that the operator ${\bm a}_i^{\pm}$
increases (decreases) the index $a_i$ by one.

Similar integration-by-parts relations for lattice integrals were
first studied in \cite{Caracciolo:1991cp,Burgio:1996ji}, where it was
shown that the entire class of integrals $H(\{a_k\};n)$ can be reduced
to $d$ master integrals in $d$ dimensions. Here, we neither attempt to
solve these equations explicitly nor to rewrite them in such a form
that the reduction of a given index is manifest. Instead, we adopt a
brute force strategy and use computer algebra to explicitly solve the
equations for a given range of indices.  An efficient algorithm for
solving such recurrence relations has been described in
\cite{Laporta:2001dd}. First, a criterion which selects a simpler
integral out of any two integrals is chosen. Typically, integrals with
lower values of the indices are considered to be simpler. The above
equations are then solved for a very limited range of indices, using
Gauss's elimination method. The calculation is repeated after
supplementing the chosen set of equations with a few relations
involving higher index values. By iterating this procedure, the
equations (\ref{eq:scalarRel}) can be solved for the entire index
range needed in a given calculation.  The advantage of this brute
force method is that it immediately generalizes to integrals involving
more complicated propagators (e.g.~those of Wilson fermions) or to
higher loops.

It is possible and advantageous to choose very convergent master
integrals, e.g. the integrals
\begin{equation}
H(\{\vec{1}\},-n)=
\int_{-\pi}^{\pi} d^4 k\, \left(\frac{{\widehat{k}}^2}{4}\right)^{n-\delta}
\end{equation}
with $n=0,1,2,3$. The expansion of these integrals in $\delta$ is then
obtained by expanding the integrand and it is trivial to numerically
evaluate them to arbitrary precision. The tadpole integral with two
powers of the boson propagator, for example, is
\begin{multline*}
H(\{\vec{1}\},2)=\left(- \frac{3}{{\delta }^2} +
     \frac{703}{72\,\delta } -\frac{2795}{144} \right)
     \,H(\{\vec{1}\},0) \\ 
+ \left( \frac{21}{2\,{\delta }^2} -
     \frac{3763}{144\,\delta }+ \frac{13313}{288} \right)
     \,H(\{\vec{1}\},-1) \\ 
+ \left( - \frac{65}{8\,{\delta }^2} + \frac{4993}{288\,\delta }-\frac{16801}{576} \right) \,H(\{\vec{1}\},-2) \\ 
+ \left( \frac{27}{16\,{\delta
     }^2} - \frac{207}{64\,\delta }+\frac{681}{128} \right)
     \,H(\{\vec{1}\},-3) +{\cal O}(\delta)\,.
\end{multline*}
Note that the $1/\delta^2$-pole is spurious: its coefficient vanishes.

The result for the hard part of the massive tadpole in Eq.~(\ref{eq2}) has the form
\begin{multline}
G^{\rm hard}(m)={b_1} + m^2\left( -{b_2} + 
\frac{1}{16\pi^2}\Big(\frac{1}{\delta} + \ln 4\Big)\!\right)\nonumber\\ + m^4\left({b_3}-\frac{1}{128\pi^2}\Big(\frac{1}{\delta}-\ln 4\Big)\!\right) +{\cal O}(m^6)\, .
\end{multline}
Note that the hard part is an analytic function of $m$. All
nonanalytic dependence on masses and momenta arises in the soft
part. As mentioned earlier, the $1/\delta$-poles cancel between the
hard and the soft part. The constants $b_1$, $b_2$, $b_3$ are obtained
by numerically evaluating the master integrals. It turns out that
these three constants are sufficient to express the hard part of any
bosonic one-loop integral to any order in the expansion around the continuum limit. Their numerical values are given in \cite{Becher:2002if}. 

\section{Staggered fermions\label{sec:staggered}}

The strategy outlined in the previous two sections did not rely on the
specific form of the propagator. In \cite{Becher:2002if}, we have
applied the technique to HQET and to QCD with Wilson fermions and
calculated a number of QCD one-loop self-energies. The recursion
relations between the hard integrals are more complicated than in the
bosonic case and the number of master integrals is larger. There are seven
master integrals for HQET and sixteen for QCD with Wilson fermions.

We now show how to use the method to calculate loop integrals for
staggered fermions, which are a common choice to put fermions on
the lattice.  The staggered fermion action is obtained by reducing the
number of components of the fermion field after spin diagonalizing the
naive lattice fermion action. In perturbative calculations it is
convenient to work with the naive fermion action and perform the
staggering only at the end of the calculation.  The fermion doubling
inherent in the naive discretization of the fermion action manifests
itself in the appearance of multiple soft regions. The sixteen zeros
of the propagator denominator of naive fermions,
\begin{equation}
D_F=\sum_\mu \frac{1}{4}\sin^2 k_\mu
=\sum_\mu \frac{\eta_i^2}{(1+\eta_i^2)^2}\, ,
\end{equation}
in the Brillouin zone give rise to sixteen propagating
fermions. Correspondingly loop integrals which involve naive fermion
propagators have sixteen soft regions. One of those is the region
where all components of the loop momentum $\eta_i$ are small. The
fifteen additional doubler contributions arise after transforming one
or several components of the integration momentum as
$\eta_i\rightarrow 1/\eta_i$ and then expanding the resulting
integrand around small $\eta_i$. For purely fermionic integrals, the
sixteen soft contributions are all equal. In integrals with both boson
and fermion propagators, the boson propagator is far off-shell for the
doubler contributions and shrinks to a point upon expanding.

The contribution of the hard part is obtained as usual, by expanding the loop
integral in external momenta and particle masses. The lattice tadpole integrals
that occur in the case at hand are
\begin{equation*}
H(\{a_i\},n,m)=\prod_{i=1}^4\int \frac{d\eta_i}{(1+\eta_i^2)^{a_i}}\,\frac{1}{D_B^n}\,\frac{1}{D_F^{m+\delta}}.
\end{equation*}
The regulator has to be on the fermion propagator in order to regulate both the singularities at $\eta_\mu=0$ {\em and} $\eta_\mu=\infty$. The integration-by-parts and partial fractioning relations for this class of integrals are
\begin{eqnarray}
0&=&\bigg\{1 + 2\, a_i\,({\bm a}_i-1) + 2\,{\bm a}_i\, ({\bm a}_i-1)\, n\,{\bm n} \nonumber\\
&&+ 2\,{\bm a}_i\, (\,{\bm a}_i-1)\, (2\,{\bm a}_i-1)\, (m+\delta) \,{\bm m}\bigg\}\,H\, ,  \nonumber\\ 
&& \nonumber\\ 
0&=&\bigg\{1-  \sum_{i=1}^{d}\,\,{\bm a}_i\,(1 - \,{\bm a}_i)\,{\bm m}\bigg\}\,H \, , \\ 
0&=&\bigg\{1 - \sum_{i=1}^{d} (1 - \,{\bm a}_i)\,{\bm n}\bigg\}\,H\, .\nonumber
\end{eqnarray}
These relations are sufficient to reduce all integrals $H$ to sixteen
convergent master integrals. 

\section{Summary and conclusion}

We have applied methods developed for the evaluation of continuum loop
integrals to calculations in lattice perturbation theory. The
technique of asymptotic expansions is used to expand lattice loop
integrals around the continuum limit. After performing the  expansion,
all nontrivial dependence on momenta and masses is encoded in
continuum loop integrals. The only genuine lattice integrals left are
massless tadpole integrals. With the help of integration-by-part
relations we then reduce all tadpole integrals to a small number of
master integrals. Except for the numerical evaluation of the master
integrals, the entire calculation is performed analytically. 

The technique we have presented does not depend on a specific form of the
lattice action and after illustrating it with a simple bosonic
integral, we have discussed how to apply it to the case of staggered
fermions. Since the techniques we have been using were
developed for multi-loop integrals, we hope that this method will also
be useful beyond one loop.

This research was supported by the DOE under grants
number DE-AC03-76SF00515 and DE-FG02-97ER41022.

\end{document}